\documentclass[twocolumn,english,superscriptaddress]{revtex4}
\usepackage[T1]{fontenc}
\usepackage[latin9]{inputenc}
\usepackage{amsmath}
\usepackage{amssymb}
\usepackage{stmaryrd}
\usepackage{graphicx}
\usepackage{esint}

\makeatletter

\newcommand*\LyXThinSpace{\,\hspace{0pt}}

\@ifundefined{textcolor}{}
{%
 \definecolor{BLACK}{gray}{0}
 \definecolor{WHITE}{gray}{1}
 \definecolor{RED}{rgb}{1,0,0}
 \definecolor{GREEN}{rgb}{0,1,0}
 \definecolor{BLUE}{rgb}{0,0,1}
 \definecolor{CYAN}{cmyk}{1,0,0,0}
 \definecolor{MAGENTA}{cmyk}{0,1,0,0}
 \definecolor{YELLOW}{cmyk}{0,0,1,0}
}

\usepackage{babel}

\makeatother

\usepackage{babel}
\begin{document}

\title{Unconventional multi-band superconductivity in bulk SrTiO$_{3}$
and LaAlO$_{3}$/SrTiO$_{3}$ interfaces}

\author{Thaís V. Trevisan}

\affiliation{School of Physics and Astronomy, University of Minnesota, Minneapolis
55455, USA}

\affiliation{Instituto de Física Gleb Wataghin, Unicamp, Rua Sérgio Buarque de
Holanda, 777, CEP 13083-859 Campinas, SP, Brazil}

\author{Michael Schütt}

\affiliation{School of Physics and Astronomy, University of Minnesota, Minneapolis
55455, USA}

\author{Rafael M. Fernandes}

\affiliation{School of Physics and Astronomy, University of Minnesota, Minneapolis
55455, USA}
\begin{abstract}
Although discovered many decades ago, superconductivity in doped SrTiO$_{3}$
remains a topic of intense research. Recent experiments revealed that,
upon increasing the carrier concentration, multiple bands cross the
Fermi level, signaling the onset of Lifshitz transitions. Interestingly,
$T_{c}$ was observed to be suppressed across the Lifshitz transition
of oxygen-deficient SrTiO$_{3}$; a similar behavior was also observed
in gated LaAlO$_{3}$/SrTiO$_{3}$ interfaces. Such a behavior is
difficult to explain in the clean theory of two-band superconductivity,
as the additional electronic states provided by the second band should
enhance $T_{c}$. Here, we show that this unexpected behavior can
be explained by the strong pair-breaking effect promoted by disorder,
which takes place if the inter-band pairing interaction is subleading
and repulsive. A consequence of this scenario is that, upon moving
away from the Lifshitz transition, the two-band superconducting state
changes from opposite-sign gaps to same-sign gaps. 
\end{abstract}
\maketitle
The elucidation of the origin of the superconductivity of doped SrTiO$_{3}$
(STO) remains a widely debated issue, with several proposals aiming
to explain why such a dilute system with a very small Fermi energy
becomes a superconductor. Indeed, the pairing glue has been attributed
to mechanisms as diverse as localized phonons \cite{Gorkov16}, ferroelectric
fluctuations \cite{Lonzarich14,Balatsky_QCP,Gamboa18}, and plasmons \cite{PLee16}.
Experimentally, it is well established that the superconducting (SC)
transition temperature $T_{c}$ displays a dome-like shape as the
carrier concentration $n$ is increased by oxygen reduction or niobium
doping \cite{Schooley64}. 

Recently, quantum oscillation measurements \cite{Behnia_fermi_surfaces}
revealed that multiple electron-like bands located at the center of
the tetragonal Brillouin zone cross the Fermi level at the same concentrations
$n$ where SC is observed. This implies that STO is a multi-band superconductor,
in agreement with earlier reports of multiple SC gaps in the tunneling
spectrum of doped STO \cite{Binnig80}. Combined with thermal conductivity
data below $T_{c}$ that indicates nodeless gaps \cite{Behnia_nodeless},
this set of data suggests that bulk STO is characterized by a multi-$s$-wave
gap structure \cite{Fernandes13,Balatsky_2bands}. These works, however,
did not resolve the relative sign of the gaps on different bands.
It generally encodes important information about the microscopic pairing
interaction, since same-sign gaps (called $s^{++}$ pairing) generally
arise from an attractive inter-band interaction, whereas opposite-sign
gaps (called $s^{+-}$ pairing) is usually the result of repulsion.

\begin{figure}
\includegraphics[width=1\columnwidth]{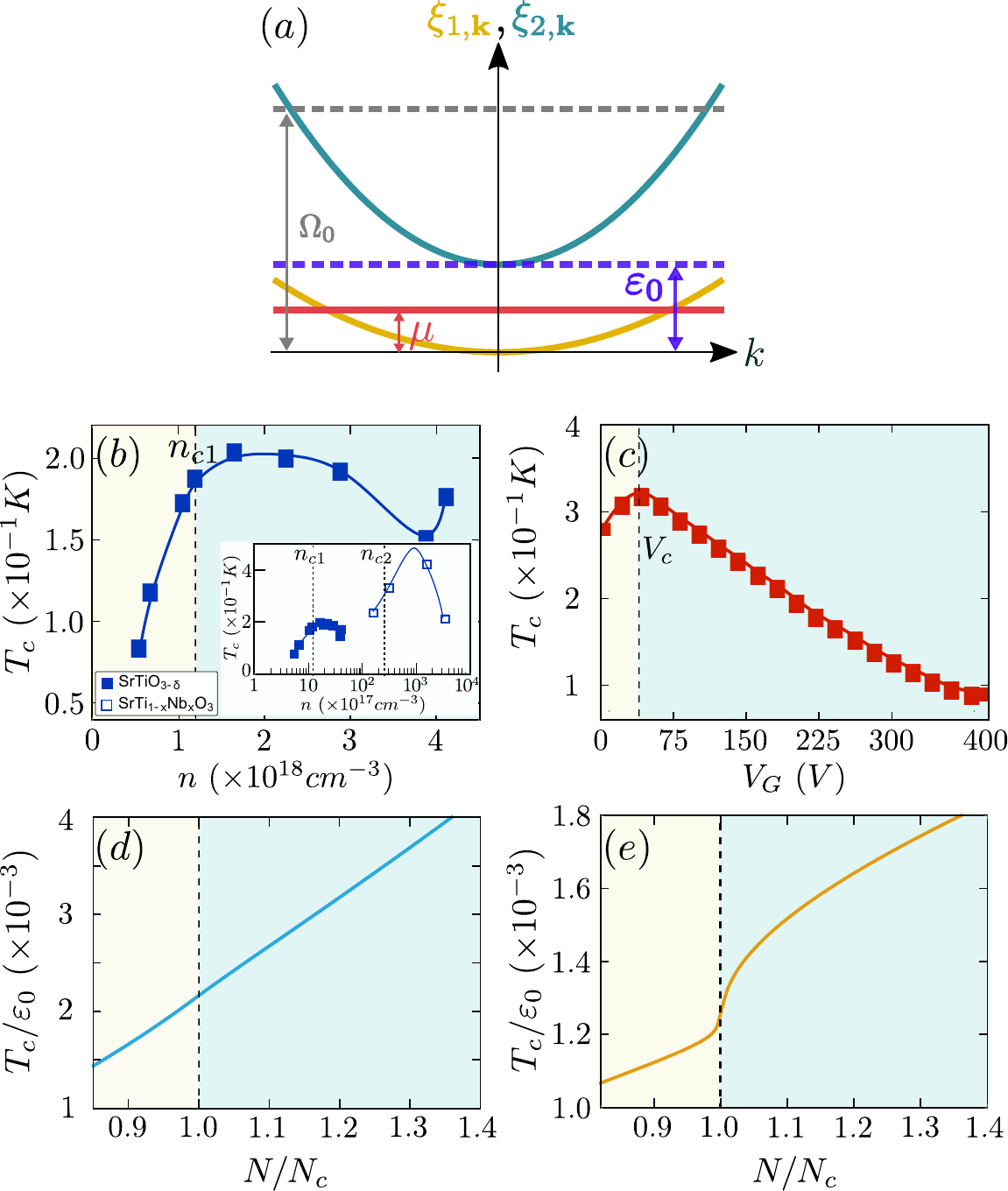} \caption{(color online) (a) Illustration of the two-band model used in this
work, where $\varepsilon_{0}$ is the energy separation between the
bottom of the two electron-like bands, $\Omega_{0}\gg \varepsilon_0$ is the interaction
cutoff, and $\mu$ is the tunable
chemical potential. Panels (b) and (c) reproduce the experimental
$T_{c}$ reported in Refs. \cite{Behnia_fermi_surfaces} and \cite{Ilani12},
respectively. In bulk STO (b), the carrier concentration $n$ is changed
by chemical doping, and Lifshitz transitions take place at the critical
values $n_{c1}$ and $n_{c2}$ (see inset). In LAO/STO (c), the occupation
number is controlled by the gate voltage $V_{G}$ and a Lifshitz transition
happens at $V_{c}$. Panels (d) and (e) show the calculated $T_{c}$ as function of the number of electrons $N$ for a clean
two-band superconductor with (d) 3D bands and (e) 2D bands. The parameters are the same as in Figs. \ref{fig_2D} and \ref{fig_3D}. $N_{c}$
indicates the Lifshitz transition. \label{model}}
\end{figure}

The multi-band nature of the SC state of STO is directly manifested
in the $n$ dependence of $T_{c}$, as shown in Fig. \ref{model}(b),
which reproduces the experimental results reported in Ref. \cite{Behnia_fermi_surfaces}.
Below a doping concentration $n_{c1}$, only one band crosses the
Fermi level, and $T_{c}$ increases as $n$ increases. As the second
band crosses the Fermi level at $n=n_{c1}$, signaling a Lifshitz
transition, $T_{c}$ changes its behavior and starts decreasing for
increasing $n$. Upon increasing the concentration even further, $T_{c}$
again increases as function of $n$, reaching its maximum value inside
the region where a third band crosses the Fermi level ($n>n_{c2}$,
see inset). However, contrary to the observed behavior of $T_{c}$
near $n_{c1}$, one expects $T_{c}$ to be enhanced across a Lifshitz
transition, since additional electronic states become available for
the pairing condensate after an additional band crosses the Fermi
level. As shown in Figs.\ref{model}(d) and (e), explicit solutions
of the two-band BCS gap equations across the Lifshitz transition find
precisely such an enhancement \cite{Bianconi10,Fernandes10,Fernandes13,Grilli13,Bang14,Hirschfeld15,Hirschfeld16,Chubukov16,Valentinis16},
regardless of whether the SC state is $s^{++}$ or $s^{+-}$.

Interestingly, such a suppression of $T_{c}$ across a Lifshitz transition
is also observed in gated LaAlO$_{3}$/SrTiO$_{3}$ (LAO/STO) interfaces
\cite{Mannhart07,Ilani12}, which have also been proposed to be multi-band
superconductors \cite{Grilli13,Schmalian15}. As shown in Fig. \ref{model}(c),
which reproduces the experimental results of Ref. \cite{Ilani12},
the maximum value of $T_{c}$ happens very close to the critical gate
voltage $V_{c}$ for which an additional band crosses the Fermi level.
This observation is yet another similarity between the SC properties
of bulk STO and LAO/STO interfaces. Previous works have indeed proposed
that the microscopic mechanism behind the LAO/STO interfacial superconductivity
may be the same as that of bulk STO \cite{vanderMarel16}. Thus, elucidating
the counter-intuitive behavior of $T_{c}$ across the Lifshitz transitions
in both materials is fundamental to advance our understanding of the
SC in these systems.

In this paper, we show that the suppression of $T_{c}$ across a Lifshitz
transition can be naturally explained if the system is \emph{dirty}
and has dominant \emph{attractive intra-band interactions} combined
with subleading \emph{repulsive inter-band interactions}. The reason
for the suppression follows from the interplay of two opposing effects.
On the one hand, an increase of $T_{c}$ is favored by the fact that
additional states become part of the SC condensate when the second
band crosses the Fermi level. On the other hand, because the inter-band
interaction is repulsive, inter-band impurity scattering is strongly
pair-breaking, favoring a suppression of $T_{c}$ when both bands
cross the Fermi level. For sufficiently large impurity scattering,
the latter effect wins, and $T_{c}$ is suppressed. We propose that
this is the case for both bulk STO and LAO/STO interfaces. Importantly,
we find that the character of the multi-band SC state across the Lifshitz
transition changes from $s^{+-}$ to $s^{++}$. We argue that this $s^{+-}$ to $s^{++}$ crossover may be the reason why certain probes only seem to identify one gap in the regime where multiple bands are known to cross the Fermi level \cite{Mannhart17}.

Our general microscopic model consists of two concentric electron-like
bands with parabolic dispersions $\xi_{1,\mathbf{k}}=\frac{k^{2}}{2m_{1}}-\mu$
and $\xi_{2,\mathbf{k}}=\varepsilon_{0}+\frac{k^{2}}{2m_{2}}-\mu$,
similarly to Ref. \cite{Fernandes13}. We note that microscopic details such as the anisotropic effective mass of bulk STO and the Rashba spin-orbit coupling of LAO/STO do not affect our main conclusions \cite{Schimalian15,Marel14}. Here, $k$ is the momentum,
$\mu>0$ is the chemical potential, $m_{i}$ are the effective band
masses, and $\varepsilon_{0}>0$ is the energy difference between
the bottom of the two bands (see Fig. \ref{model}(a)). Upon increasing
$\mu$, band $2$ crosses the Fermi level when $\mu=\varepsilon_{0}$,
signaling a Lifshitz transition. The pairing interaction coupling
bands $i$ and $j$ is denoted by $V_{ij}$. Although their origin
is not specified here, we allow the interactions to be attractive
($V_{ij}<0$) or repulsive ($V_{ij}>0$). Based on the experimental
observations indicating nodeless gaps in STO \cite{Behnia_nodeless},
we consider $V_{ij}$ to be momentum-independent. As a result, there
are two isotropic gaps $\Delta_{1}$ and $\Delta_{2}$ in bands $1$
and $2$, respectively.

The pairing interaction extends to an energy cutoff $\Omega_{0}$,
which is assumed to be larger than the typical Fermi energy scale
$\varepsilon_{0}$ of the model \cite{Mazin11}. Although $\Omega_{0}>\varepsilon_{0}$,
we can still neglect vertex corrections and write down the linearized
BCS-like gap equations as long as $\left|\lambda_{ij}\right|\ll1$.
Here, $\lambda_{ij}\equiv-\rho_{j,0}V_{ij}$ is the dimensionless
pairing interaction, with $\rho_{j,0}$ denoting the density of states
of band $j$ at the energy $\varepsilon_{0}$ above the bottom of
the band. Note that, in this notation, $\lambda_{ij}>0$ ($\lambda_{ij}<0$)
implies attractive (repulsive) interaction. The gap equation then
becomes a $2\times2$ eigenvalue equation of the form:
\begin{equation}
\left(\begin{array}{c}
\Delta_{1}\\
\Delta_{2}
\end{array}\right)=\left(\begin{array}{cc}
\lambda_{11} & \lambda_{12}\\
\lambda_{21} & \lambda_{22}
\end{array}\right)\hat{A}_{\mathrm{clean}}\left(\mu,T_{c}\right)\left(\begin{array}{c}
\Delta_{1}\\
\Delta_{2}
\end{array}\right)\text{.}\label{gap_eq}
\end{equation}
where $\hat{A}_{\mathrm{clean}}\left(\mu,T_{c}\right)$ is a diagonal
matrix with matrix elements $(\hat{A}_{\mathrm{clean}})_{ii}=\pi T_{c}\sum_{n}f_{n,i}$. Here, we introduced $f_{n,i}\equiv\left\langle (\omega_{n}^{2}+\xi^{2})^{-1}\right\rangle _{i}^{\Omega_{0}}$
and the notation $\left\langle \mathcal{O}(\xi)\right\rangle _{i}^{\Omega_{0}}\equiv(1/\pi\rho_{i,0})\int_{W_{i}}^{\Omega_{0}}d\xi\rho_{i}(\xi)\mathcal{O}(\xi)$,
with $W_{i}$ denoting the bottom of band $i$, i.e. $W_{1}=-\mu$
and $W_{2}=-\mu+\varepsilon_{0}$. Of course, if $\left|W_{i}\right|\gg\Omega_{0}$,
the problem reduces to the standard two-band BCS problem, with $(\hat{A}_{\mathrm{clean}})_{ii}=(\rho_{i,F}/\rho_{i,0})\ln\left(1.13\Omega_{0}/T_{c}\right)$,
and $\rho_{i,F}$ denoting the density of states at the Fermi level.
Here, however, we are interested in the evolution of $T_{c}$ across
the Lifshitz transition, with $\left|W_{i}\right|\ll\Omega_{0}$.
Diagonalization of Eq. (\ref{gap_eq}) gives $T_{c}\left(\mu\right)$.
Note that $\mu$ is not the $T=0$ chemical potential, but its value
at $T_{c}$. Since the Fermi energy of one of the bands is small near
the Lifshitz transition, these two values may be different \cite{Chubukov16}.
For this reason, it is more meaningful to express $T_{c}\left(N\right)$,
where $N$ is the particle number.

To solve the gap equations, we need to set the values of the interactions.
Since $T_{c}$ is of the same order of magnitude on either side of
the Lifshitz transition (see Fig. \ref{model}(b) and (c)), the SC
state should be dominated by intra-band pairing. Microscopically,
this can be motivated from the facts that each band consists majoritarily
of a different $t_{2g}$ Ti orbital ($3d_{xz}$, $3d_{yz}$, $3d_{xy}$)
\cite{Mazin11}, and that intra-orbital pairing is presumably favored
over inter-orbital pairing. In this case, where $\left|\lambda_{i=j}\right|\gg\left|\lambda_{i\neq j}\right|$,
SC is achieved only with attractive intra-band interaction. The inter-band
interaction, however, can be either attractive or repulsive. Furthermore,
since $T_{c}\ll\varepsilon_{0}$ in STO, we consider $\left|\lambda_{ii}\right|\ll1$.
For concreteness, in what follows, we set $\lambda_{11}=\lambda_{22}=0.13$
and $\lambda_{12}=\lambda_{21}$, with $|\lambda_{12}|=\lambda_{11}/10$.
We also set the pairing interaction energy scale $\Omega_{0}/\varepsilon_{0}=5$,
the band width $\Lambda=\Omega_{0}$ and the density of states of
the two bands to be the same, $\rho_{1,0}=\rho_{2,0}$. The results
reported here do not change significantly for other choices of parameters
that respect the hierarchy of energy scales discussed above \cite{Trevisan18}.

\begin{figure}
\centering \includegraphics[width=0.82\linewidth]{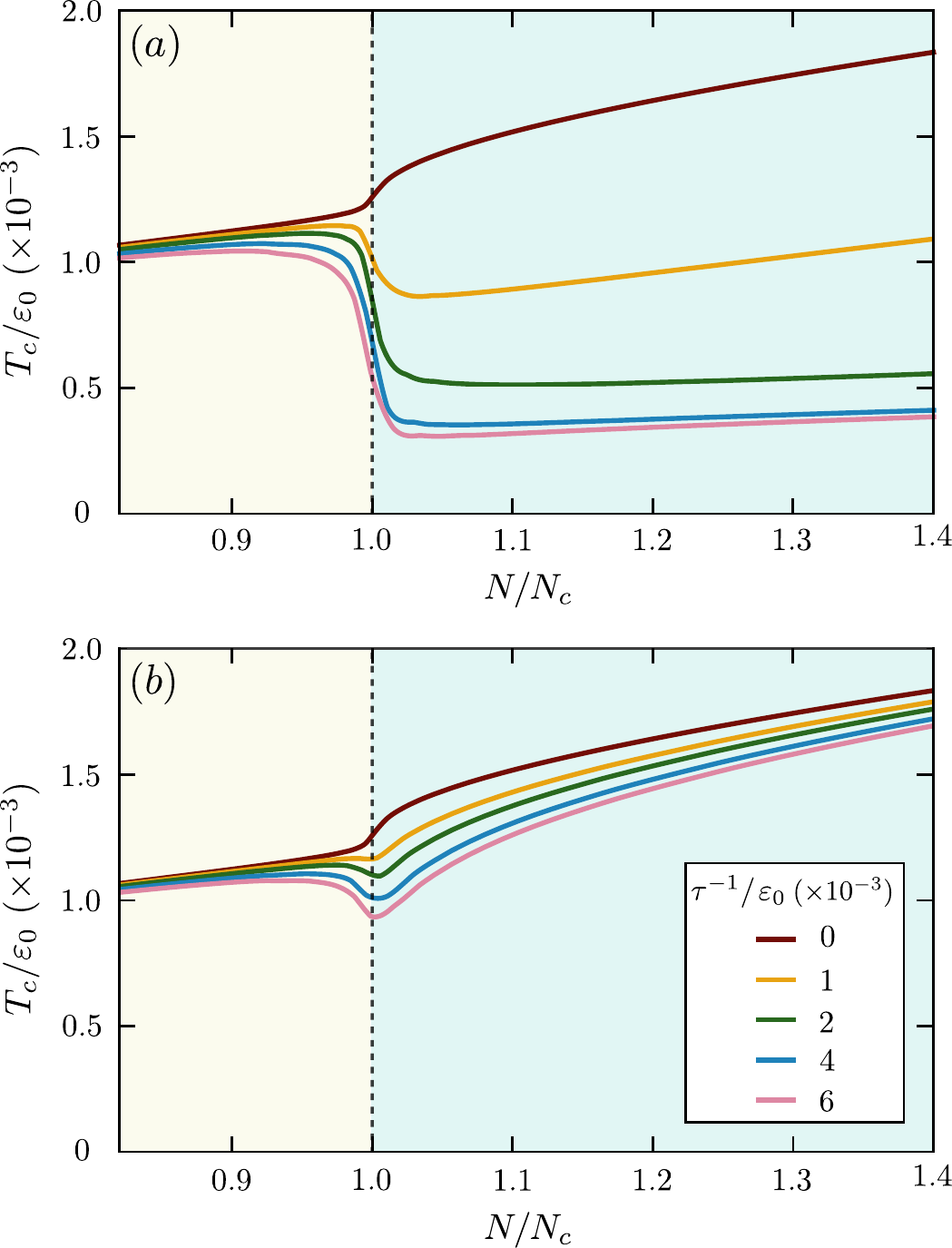} \caption{(color online) SC transition temperature $T_{c}$ as function of the
occupation number $N$ for 2D bands. A Lifshitz transition takes place
at $N_{c}$. Different point-like impurity scattering rates $\tau^{-1}$
are shown for an inter-band pairing interaction $\lambda_{12}$ that
is either (a) repulsive or (b) attractive. \label{fig_2D}}
\end{figure}

In Figs.\ref{model}(d) and (e), we show the calculated behavior of
$T_{c}$ across the Lifshitz transition, which takes place at the
critical concentration $N_{c}$. Regardless of whether the inter-band
pairing is repulsive or attractive, there is always an enhancement
of $T_{c}$ at the Lifshitz transition, which can be sharper or milder
depending on the dimensionality. This result is in agreement with
several previous analysis of similar models \cite{Fernandes10,Fernandes13,Grilli13,Bang14,Hirschfeld15,Hirschfeld16,Chubukov16},
but in sharp disagreement with the experimental behavior of $T_{c}(n)$
observed in STO (near $n_{c1}$) and LAO/STO (near $V_{c}$).

A key ingredient missing in our model is disorder. From the STO residual
resistivity data of Refs. \cite{Behnia_fermi_surfaces,Behnia_science},
an estimate of the scattering rate gives $\tau^{-1}\sim10\,T_{c}$.
At first sight, one may think that disorder does not impact SC due
to Anderson's theorem, since the gap is s-wave. However, Anderson's
theorem applies for a single band with a fully isotropic gap. When the gaps from different bands are not equal, $T_{c}$ is suppressed
by disorder \cite{Golubov97}. This effect is much more significant
when the inter-band interaction is repulsive, as in this case inter-band
impurity scattering becomes strongly pair-breaking. This insight offers
a plausible scenario for the suppression of $T_{c}$ across a Lifshitz
transition: as long as the inter-band pairing is repulsive, the detrimental
effects of inter-band scattering are amplified when the second band
crosses the Fermi level. These effects are however tamed when pairing
is predominantly confined to a single band on the other side of the
Lifshitz transition. Note that, in contrast to the case discussed in this paper, if the impurity is magnetic, it is pair-breaking for both repulsive and attractive inter-band interactions.

To quantitatively assess the suitability of this scenario, we include
intra-band and inter-band disorder potentials $v$ and $u$, respectively,
in our two-band model via the self-consistent Born approximation (see
also Ref. \cite{Hirschfeld16}). The only change in the gap equation
(\ref{gap_eq}) is the new matrix $\hat{A}_{\mathrm{dirty}}$ that
replaces its clean counterpart $\hat{A}_{\mathrm{clean}}$. Defining
$\tilde{f}_{n,i}=\left\langle \left[\tilde{\omega}_{n,i}^{2}+\left(\xi+h_{n,i}\right)^{2}\right]^{-1}\right\rangle _{i}^{\Omega_{0}}$,
the matrix elements are given by $(\hat{A}_{\mathrm{dirty}})_{ij}=\pi T_{c}\sum_{n}\tilde{f}_{n,i}M_{ij}/\mathcal{D}_{n}$,
with:

\begin{equation}
M_{ij}=\left(1-\frac{\tau_{\bar{i}\bar{i}}^{-1}}{2}\tilde{f}_{n,\bar{i}}\right)\delta_{i,j}+\left(\frac{\tau_{ij}^{-1}}{2}\tilde{f}_{n,j}\right)\left(1-\delta_{i,j}\right)
\end{equation}

Here, $\bar{i}=1,2$ for $i=2,1$ and $\mathcal{D}_{n}=\left(1-\frac{\tau_{11}^{-1}}{2}\tilde{f}_{n,1}\right)\left(1-\frac{\tau_{22}^{-1}}{2}\tilde{f}_{n,2}\right)-\frac{\tau_{12}^{-1}\tau_{21}^{-1}}{4}\tilde{f}_{n,1}\tilde{f}_{n,2}$.
Furthermore, $\tau_{ij}^{-1}=2\pi n_{\mathrm{imp}}\rho_{j,0}\left(\left|v\right|^{2}\delta_{ij}+\left|u\right|^{2}\delta_{\bar{i}j}\right)$
are the scattering rates, and $n_{\mathrm{imp}}$ is the impurity
concentration. Finally, $\tilde{\omega}_{n,i}=\omega_{n}+\sum_{j}\tau_{ij}^{-1}\tilde{\omega}_{n,j}\tilde{f}_{n,j}/2$
is the disorder-renormalized Matsubara frequency, $\omega_{n}=\left(2n+1\right)\pi T$,
and $h_{n,i}=-\frac{1}{2}\sum_{j}\tau_{ij}^{-1}\left\langle \frac{\xi+h_{n,j}}{\tilde{\omega}_{n,j}^{2}+\left(\xi+h_{n,j}\right)^{2}}\right\rangle _{j}^{\Omega_{0}}$
is the renormalization of the band dispersions by disorder.

To solve the gap equation across the Lifshitz transition, we simultaneously
solve the self-consistent equations for $\tilde{\omega}_{n,i}$ and
$h_{n,i}$ numerically. For concreteness, we consider the case of
point-like impurities, for which $\tau_{ij}^{-1}=\tau^{-1}$ (i.e.
equal intra-band and inter-band scattering). Although here we assumed a constant $\tau^{-1}$, it is expected to increase with $N$. While this will not affect the $T_c$ behavior in the vicinities of the Lifshitz transition, it may lead to a stronger suppression of superconductivity for larger $N$. 

\begin{figure}[t!]
\centering \includegraphics[width=0.82\linewidth]{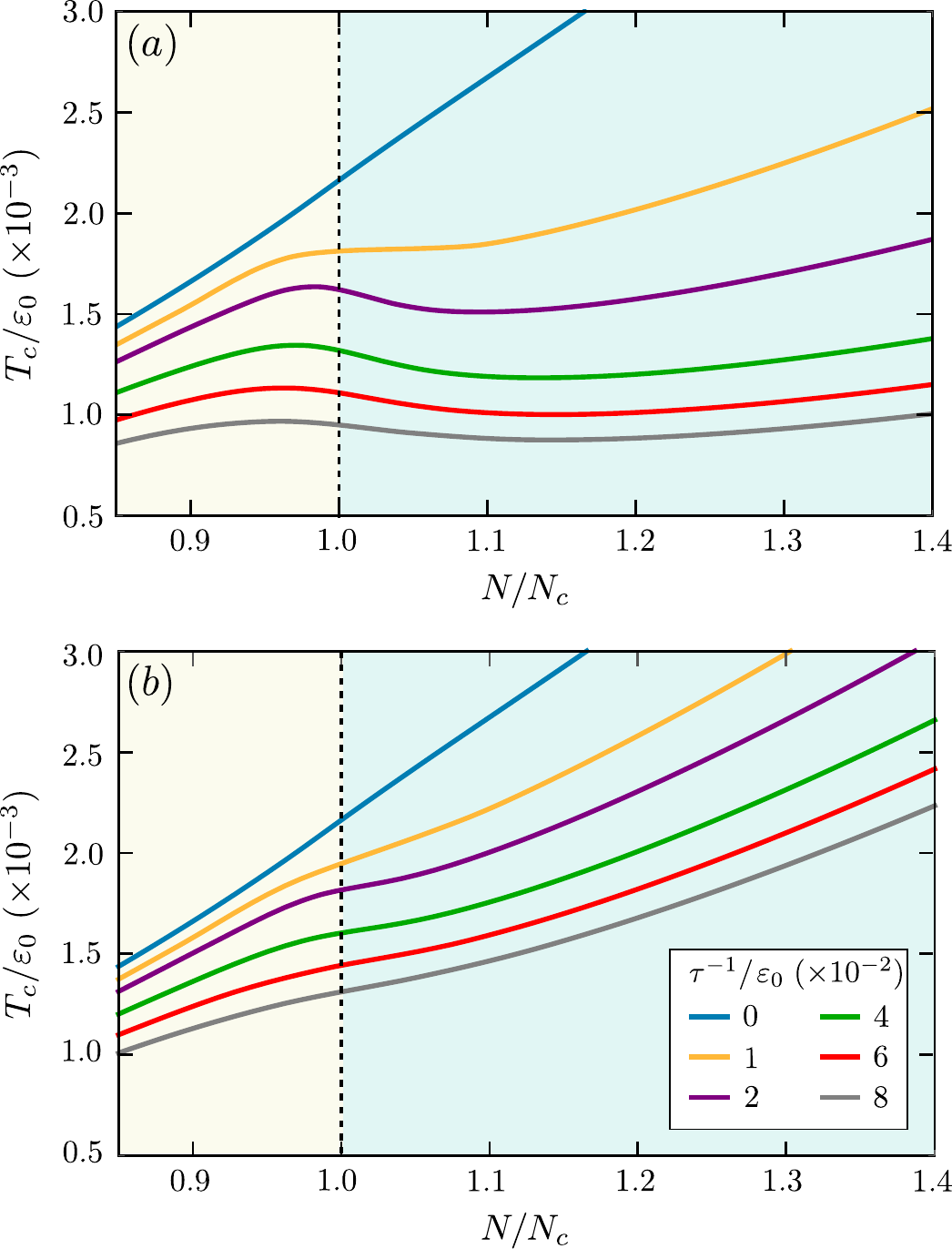} \caption{(color online) SC transition temperature $T_{c}$ as function of the
occupation number $N$ for 3D bands for different scattering rates
$\tau^{-1}$. Similarly to Fig. \ref{fig_2D}, a Lifshitz transition
takes place at $N_{c}$. Both repulsive (a) and attractive (b) inter-band
pairing interactions $\lambda_{12}$ are shown. \label{fig_3D}}
\end{figure}

We first discuss the results for 2D bands, relevant for thin films
of STO and LAO/STO interfaces. The theoretical results should thus be compared to the experimental phase diagram of Fig. \ref{model}(c). As shown in Fig. \ref{fig_2D}(a),
in the case of repulsive inter-band pairing interaction ($\lambda_{12}<0$),
increasing the impurity scattering $\tau^{-1}$ to values comparable
to the SC transition temperature of the clean system leads to a qualitative
change in the behavior of $T_{c}\left(N\right)$ across the Lifshitz
transition. Remarkably, instead of the enhancement of $T_{c}$ at
the critical concentration $N_{c}$ observed in the clean case (see
Fig. \ref{model}(e)), now the system displays a suppression of $T_{c}$
around $N_{c}$. We verified that such suppression of $T_c$ persists even when $\rho_{1,0}$ and $\rho_{2,0}$ are not equivalent. Moreover, while on the single-band side of the Lifshitz
transition $T_{c}$ is only mildly affected by disorder, on the two-band
side $T_{c}$ is more severely suppressed by increasing $\tau^{-1}$. 
This behavior is very different than the case of attractive inter-band
pairing interaction ($\lambda_{12}>0$), shown in Fig. \ref{fig_2D}(b):
not only is the $T_{c}$ suppression at $N_{c}$ much milder, but
$T_{c}$ is only weakly affected by disorder in both sides of the
Lifshitz transition. Similar results hold for the case of 3D bands,
which are relevant for doped bulk STO, whose experimental phase diagram is shown in Fig. \ref{model}(b). As shown in Fig. \ref{fig_3D}, for $\lambda_{12}<0$, the suppression of $T_{c}$ around the Lifshitz transition in the presence of strong disorder is less sharp than in the 2D case due to the fact that the density of states in 3D vanishes smoothly at the band edge.

\begin{figure}
\centering \includegraphics[width=0.82\columnwidth]{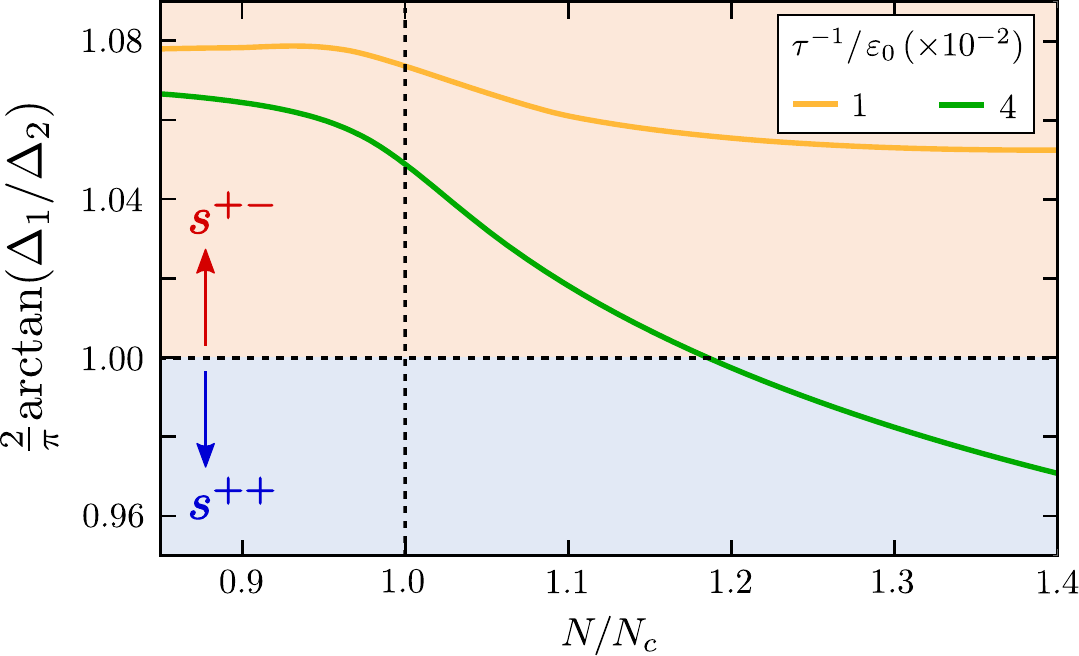}
\caption{(color online) Ratio between the two isotropic SC gaps in bands $1$
and $2$ ($\Delta_{1}$ and $\Delta_{2}$, respectively) across the
Lifshitz transition at $N=N_{c}$. These plots refer to the 3D bands
case shown in Fig. \ref{fig_3D}. For sufficiently large impurity
scattering rate, the relative sign of the two SC gaps change for $N>N_{c}$,
signaling a crossover from an $s^{+-}$ SC state to an $s^{++}$
one. \label{fig_splusplus}}
\end{figure}

In both Figs. \ref{fig_2D}(a) and \ref{fig_3D}(a), $T_{c}$ seems
to recover far enough away from the Lifshitz transition, suggesting that SC is able to counteract the detrimental pair-breaking
effect of inter-band impurity scattering, and become less sensitive to disorder, even though the inter-band pairing interaction is always repulsive ($\lambda_{12}<0$). To shed light on this behavior, in Fig. \ref{fig_splusplus} we plot $\mathrm{arctan}\left(\Delta_{1}/\Delta_{2}\right)$ as function of $N$ for the case of 3D bands (a similar behavior holds in 2D).
For strong enough disorder strength, we note that the relative sign
of the two gaps changes, as the system moves away from the Lifshitz
transition, signaling an $s^{+-}$ to $s^{++}$ crossover. Previously,
such a crossover was shown to take place as function of increasing
disorder \cite{Efremov11}; here, we find it to be tuned by the chemical
potential for fixed disorder strength.

In summary, we showed that the presence of impurity scattering dramatically
changes the behavior of $T_{c}$ across a two-band Lifshitz transition,
provided that the SC state is dominated by intra-band pairing interaction
and that the subleading inter-band pairing interaction is repulsive.
This is a consequence of two opposing effects that take place close to the Lifshitz transition: an enhancement of the number
of electronic states available to participate in the SC condensate,
which enhances $T_{c}$, and an enhancement of pair-breaking effects
arising from inter-band impurity scattering, which suppresses $T_{c}$.
As a result, for impurity scattering rates comparable to the SC transition
temperature of the clean system, $T_{c}(N)$ displays a maximum near
the Lifshitz transition at $N=N_{c}$, followed by a subsequent enhancement
for increasing $N>N_{c}$, which is more pronounced
in the case of 3D bands, as shown in Figs. \ref{fig_2D}(a) and \ref{fig_3D}(a).

These general results are in qualitative agreement with the experimental
SC phase diagrams of doped bulk STO and gated LAO/STO, as shown in
Fig. \ref{model}, indicating that the multi-band superconductivity
in these materials is unconventional and promoted by inter-band repulsion.
Note that, in the case of bulk STO, there are two Lifshitz transitions,
and the second one seems to have a weaker effect on $T_{c}$. While
a three-band calculation is beyond the scope of this work, it is reasonable
to expect that, when same-sign SC is well-established in two larger
Fermi surfaces, the effects of a third incipient smaller Fermi surface
will be weaker. An important prediction of our model is the $s^{+-}$
to $s^{++}$ SC crossover near the Lifshitz transition, which could be realized in the phase diagrams of bulk STO and LAO/STO. While direct verification of such a crossover is difficult, indirect effects could be detected. For instance, since the magnitude of the gap of the incipient band across the crossover first decreases, then vanishes, and finally increases again, certain experimental probes may not identify it. This
may result in spectroscopic and thermodynamic signatures that are
consistent with a single-band superconductor, despite the fact that
two bands cross the Fermi level. Interestingly, recent optical conductivity
data in doped STO were explained in terms of a single SC gap \cite{Mannhart17}. 
\begin{acknowledgments}
We thank A. Balatsky, K. Behnia, A. Chubukov, M. Gastiasoro, B. Jalan,
C. Leighton, G. Lonzarich, and V. Pribiag for fruitful discussions.
This work was primarily supported by the U.S. Department of Energy
through the University of Minnesota Center for Quantum Materials,
under award DE-SC-0016371 (R.M.F.). T.V.T. acknowledges the support
from the São Paulo Research Foundation (Fapesp, Brazil) via the BEPE scholarship. 
\end{acknowledgments}

\end{document}